\documentclass[aps,prb,amsmath,amssymb,twocolumn,longbibliography, superscriptaddress]{revtex4-2}
\usepackage{graphicx}
\usepackage{dcolumn}
\usepackage{bm}
\usepackage{mathrsfs}
\usepackage{subfigure}
\usepackage{natbib}
\usepackage{amssymb}
\usepackage{multirow}
\usepackage{float}
\usepackage{color}
\usepackage{ulem}
\usepackage{mathrsfs}

\begin{document}

\title{Surface superconductor-insulator transition induced by an electric field}

\author{Liyu Yin}
\affiliation{Key Laboratory of Optical Field Manipulation of Zhejiang Province, Department of Physics, Zhejiang Sci-Tech University, 310018 Zhejiang, China}

\author{Yunfei Bai}
\affiliation{Key Laboratory of Optical Field Manipulation of Zhejiang Province, Department of Physics, Zhejiang Sci-Tech University, 310018 Zhejiang, China}

\author{Ming Zhang}
\affiliation{Key Laboratory of Optical Field Manipulation of Zhejiang Province, Department of Physics, Zhejiang Sci-Tech University, 310018 Zhejiang, China}

\author{A. A. Shanenko}
\affiliation{HSE University, 101000 Moscow, Russia}

\author{Yajiang Chen}
\email{yjchen@zstu.edu.cn}
\affiliation{Key Laboratory of Optical Field Manipulation of Zhejiang Province, Department of Physics, Zhejiang Sci-Tech University, 310018 Zhejiang, China}

\date{\today}
\begin{abstract}
It is well-known that the electric field can induce phase transitions between superconducting, metallic and insulating states in thin-film materials due to its control of the charge carrier density. Since a similar effect on the charge carriers can also be expected for surfaces of bulk samples, here we investigate the transformation of the surface states in a superconductor under an applied screened electric field. Our study is performed by numerically solving the self-consistent Bogoliubov-de Gennes equations for the one-dimensional attractive Hubbard model. It is found that the surface insulating regime occurs at sufficiently large (but still experimentally accessible) electric fields. Our calculations yield the phase diagram of the surface superconducting, metallic, and insulating states for a wide range of temperatures and applied fields. Our results are in qualitative agreement with the phase diagram obtained by the transport measurements for (Li, Fe)OHFeSe thin flakes [Sci. Bull. 64, 653 (2019); ACS Nano 14, 7513 (2020)].
\end{abstract}
\maketitle
\section{Introduction}

Due to the capability of modulating the carrier concentration, the electric field has been utilized as one of the most important experimental tools in the field of superconductivity for several decades~\cite{glover1960,bonfiglioli1962,meissner1967,golokolenov2021,elalaily2021}. In particular, electric-field effects on the superconductor-metal transition have been revealed theoretically and experimentally. Electrostatic charging created by an external electric field ($E\approx2\time10^{-7}$ V/m) is able to cause a shift of the superconducting transition temperature ($\Delta T_c\approx10^{-5}$ K) in both tin and indium thin films~\cite{glover1960,bonfiglioli1962}. Electric fields change the energy of itinerant electrons in atomically thin flakes of NbSe$_2$, which results in shifting the chemical potential and changing the density of states in the Debye window and, in turn, in altering $T_c$~\cite{staley2009}. Though electrons are heavily affected by an electric field near surfaces, it has been shown by the electrical and thermal conductivity measurements that tin films possess no surface superconductivity in the presence of an electric field~\cite{meissner1967}. However, for systems with a sufficiently low Fermi level, the surface bound states of electrons induced by electric fields may result in the appearance of the multigap surface superconductivity~\cite{mizohata2013}. In addition, $T_c$ of oxide superconductors (e.g. $8$-nm-thick GdBa$_2$Cu$_3$O$_{7-x}$ films, Nb-doped SrTiO$_3$ films) can be tuned by sufficiently large electric fields due to dielectric breakdown~\cite{ahn1999,ahn2003,takahashi2004}. Furthermore, the electric field influences other superconducting properties related to the superconductor-metal transition, e.g. suppression of the critical supercurrent~\cite{golokolenov2021,elalaily2021,orus2021,paolucci2021,ritter2021,amoretti2022}.

Electric fields can also induce a superconducting state in insulators. For example, by increasing the gate voltage ($V_g$) from $0$ to $42.5$ V, the resistance of $10.22$-$\AA$-thick amorphous Sb film at $T=65$ mK~\cite{parendo2005} drops continuously from $22\, {\rm k\Omega}$ to $0$, which implies that the sample may undergo the insulator-metal and metal-superconductor transitions in sequence. Here, the electric field associated with the onset of superconductivity is sufficiently high (up to $4.2\times 10^{10}$ V/m), corresponding to the dielectric breakdown. It was found that the increasing of electron concentrations screens the electron-electron interactions, which produces an effective attractive potential and promotes the superconductive correlations. More particularly, for a pristine SrTiO$_3$ channel with size $15\,\mu{\rm m}\times 200\,\mu{\rm m}$~\cite{ueno2008}, the system undergoes a sharp superconducting transition with a mid-point critical temperature $T_c^{\rm mid}=0.4$ K at $V_g=3$ V according to transport measurements. In this case, the electric field is $2-3$ orders of magnitude weaker than the dielectric-breakdown field~\cite{paolucci2019}, and the sample is metallic at $T<0.1$ K and $V_g=2.50$ V because its resistance is about $20\,\Omega$. It means that this transition actually occurs from the metallic state to the superconducting state as the electric field increases. Similar transitions have also been observed in $2$-nm-thick GdBa$_2$Cu$_3$O$_{7-x}$ films~\cite{ahn2003}, atomically flat ZrNCl film~\cite{ye2010}, La$_{2-x}$Sr$_x$CuO$_4$ films~\cite{bollinger2011}, etc. 

Recently, transport measurements have revealed~\cite{ma2019, yin2020} the direct superconductor-insulator transition that occurs in thin (Li, Fe)OHFeSe flakes with $T\approx0$ and $V_g\approx5.13$ V. Its mechanism is not clear yet, as many important details, such as the differential conductance  ${\rm d}I/{\rm d}V$ and the $T$-dependent resistance, are missing. However, this is certainly an example where the electric-field effects play a crucial role.

In the present work, motivated by these experiments with thin (Li, Fe)OHFeSe flakes~\cite{ma2019, yin2020}, we investigate the transformation of the surface properties in a bulk superconductor under an applied electric field. In particular, we consider the effect of a screened electric field on the superconducting state near the edges of the system within the one-dimensional attractive Hubbard model at the half-filling level by numerically solving the self-consistent Bogoliubov-de Gennes (BdG) equations. Our study demonstrates that the direct surface superconductor-insulator transition does arise in the superconductor for sufficiently strong electric fields and low temperatures. Moreover, our findings are in qualitative agreement with the phase diagram obtained by the transport measurements for (Li, Fe)OHFeSe thin flakes~\cite{ma2019, yin2020}.

The present paper is organized as follows. In Sec.~\ref{sec2} we discuss the BdG equations for the one-dimensional attractive Hubbard model in the presence of an applied (screened) electric field. In our study the BdG equations are solved numerically, in a self-consistent manner, and the main points of this procedure are also outlined in Sec.~\ref{sec2}. In Sec.~\ref{sec3}, we consider numerical results for the pair potential and electron distribution together with the corresponding quasiparticle energies and wavefunctions. These results yield the phase diagram of the surface superconducting, metallic and insulating states versus the temperature and the electric-field strength. Finally, our main conclusions are given in Sec.~\ref{sec4}.

\section{Theoretical Formalism}
\label{sec2}

\subsection{Bogoliubov-de Gennes equations}
As we are interested in the qualitative picture of the surface-state transformations, our analysis can be  simplified by considering a one-dimensional chain of atoms in a parallel electric field. The corresponding attractive Hubbard model with the $s$-wave pairing and within the tight-binding approximation is based on the grand-canonical Hamiltonian~\cite{tanaka2000, takasan2019}:
 \begin{align}\label{hubbard}
\mathscr{H} - \mu \mathscr{N}_e  =& -\sum_{i\delta\sigma}t_\delta c^{\dagger}_{i+\delta,\sigma}c_{i\sigma} + \sum_{i\sigma}\big[V(i) -\mu\big] n_{i\sigma}  \nonumber \\ 
&  -g \sum_i n_{i\uparrow}n_{i\downarrow},
\end{align}
where $\mu$ is the chemical potential, and $\mathscr{N}_e$ is the total electron number operator, i.e. $\mathscr{N}_e=\sum_{i\sigma} n_{i\sigma}=\sum_{i\delta} c^{\dagger}_{i\sigma}c_{i\sigma}$, with $c_{i\sigma}$ ($c^{\dagger}_{i\sigma}$) the annihilation (creation) operator of an electron with spin $\sigma(=\uparrow,\downarrow)$ at the sites $i=0,...,N+1$. $t_\delta$ is the hopping rate of electrons between the sites $i$ and $i+\delta$. In the present study only the nearest neighbors are taken into account, i.e. $\delta=\pm1$ and thus, we have $t_\delta = t$. Finally, $g$ denotes the on-site attractive interaction between electrons resulting from the electron-phonon coupling, and $V(i)$ is the electrostatic energy appearing due to the presence of a screened electric field. 

Within the mean-field approximation one gets the effective Hamiltonian~\cite{gennes1966}
\begin{align}\label{Heff}
H_{\rm eff} =& -t\sum_{i\delta\sigma} c^{\dagger}_{i+\delta,\sigma}c_{i\sigma} + \sum_{i\sigma}\big[V(i) -\mu\big] n_{i\sigma}  \nonumber \\ 
& - \sum_i \big[\Delta(i)c^{\dagger}_{i\uparrow}c^{\dagger}_{i\downarrow} + \Delta^{*}(i)c_{i\downarrow}c_{i\uparrow}\big]
\end{align}
with $\Delta(i)$ the site-dependent superconducting pair potential. Diagonalizing $H_{\rm eff}$ through the generalized Bogoliubov-Valatin transformation~\cite{gennes1966}, we obtain the BdG equations~\cite{samoilenka2020, croitoru2020, chen2022, bai2023}
\begin{subequations}\label{bdg}
\begin{align}
\epsilon_\alpha u_\alpha(i) & =  \sum_{i'} H_{ii'}u_\alpha(i') + \Delta(i) v_\alpha(i) \\
\epsilon_\alpha v_\alpha(i) & =  \Delta^*(i)u_\alpha(i)  - \sum_{i'} H^*_{ii'} v_\alpha(i'),
\end{align}
\end{subequations}
where $H_{ii'}$ is the single-particle Hamiltonian and $\epsilon_\alpha$, $u_\alpha(i)$, and $v_\alpha(i)$ are the energy and wavefunctions of quasiparticles, respectively. The index $\alpha$ enumerates the quasiparticle states in the energy ascending order (only the states with the positive quasiparticle energies are taken into consideration)~\cite{shanenko2007, chen2018}. We apply the open boundary conditions, i.e. the quasiparticle wavefunctions vanish at $i=0$ and $N+1$. The Hartree-Fock potential is ignored in our study since its main effect is barely shifting the chemical potential~\cite{chen2009}. The single-particle Hamiltonian $H_{ii'}$ is of the form
\begin{equation}
H_{ii'} = -t\sum_{\delta=\pm1} \delta_{i',i+\delta}+\big[V(i)-\mu\big]\delta_{ii'},
\end{equation}
where the chemical potential $\mu$ is determined by the electron-filling level $\bar{n}_e = \sum_{i} n_e(i)/N$, where the electron distribution $n_e(i)$ is as follows
\begin{equation}\label{ne}
n_e(i) = 2\sum_{\alpha}\big[ f_\alpha |u_\alpha(i)|^2 + (1-f_\alpha)|v_\alpha(i)|^2\big],
\end{equation}
with $f_\alpha=f(\epsilon_\alpha)$ the Fermi-Dirac distribution. Below we focus on the half-filling case, i.e. $\bar{n}_e=1$. The spatial pair potential $\Delta(i)$ is related to the quasiparticle energies and wavefunctions by~\cite{gennes1964, chen2022, bai2023}
\begin{equation}\label{op}
\Delta(i) = g\sum_\alpha u_\alpha(i) v^*_\alpha(i)\big[1-2f_\alpha \big].
\end{equation}
Here the sum is over the quasiparticle states within the Debye window, i.e. $0\leq\epsilon_\alpha \leq \hbar\omega_D$, where $\omega_D$ is the Debye frequency. 

The BdG equations~(\ref{bdg}) are solved self-consistently together with Eqs.~(\ref{ne}) and (\ref{op}). First, we solve the BdG equations using some initial guess for the chemical potential $\mu$ and pair potential $\Delta(i)$. Second, based on this solution, we find the electron-filling level and the new pair potential according to Eqs.~(\ref{ne}) and (\ref{op}), respectively. Third, if the new pair potential differs significantly from the initial guess and/or the electron-filling level is lower or higher than the half-filling one, we go back to the first step, replacing the initial guess for the pair potential by its new variant and slightly changing the chemical potential. The procedure is repeated until the convergence of $\Delta(i)$ under the condition that $\bar{n}_e$ approaches the half-filling level.
 
\subsection{Screened electric field ${\bf E}(x)$ and electrostatic energy $V(x)$}

The parallel electric field is introduced by using the approach of two charge reservoirs with equal but opposite charges located at the opposite surfaces of the system~\cite{paolucci2019}. Instead of a uniform electric field appropriate to insulating materials~\cite{takasan2019,szalowski2014}, here we consider a screened electric field. This variant is relevant for the case when the bulk of the sample is metallic or superconductive. Then, following Ref.~\onlinecite{amoretti2022}, the electric field is written as
\begin{align}\label{Ei}
{\bf E}(x) & = E_0 \big[e^{-x/\lambda_{E}} + e^{-(L-x)/\lambda_{E}} \big]\hat{\bf{x}} \nonumber \\
& = 2\,E_{0}\,e^{-L/2\lambda_{E}} \cosh\big[\big(2x-L\big)/2\lambda_{E}\big]\hat{\bf{x}},
\end{align}
where $\lambda_E$ is the screening length, $E_0$ is the value of the electric field at the boundaries, $L$ is the chain length, i.e. $L=(N+1)a$ with $a$ the lattice constant, $x=(i-1)a$ is the site coordinate, and $\hat{\bf x}$ is the unit vector along the chain. 

The surface screening of the electric field in the presence of the transformation of the surface superconductive/metallic states to the surface insulating state is a rather complex problem, and the screening length is certainly sensitive to this transformation. One can expect that the screening length $\lambda_{E}$ is approximately proportional to Fermi wavelength $\lambda_{F}$ but the value of the proportionality factor can vary, depending on a particular surface state. However, our analysis demonstrates that the qualitative picture of our results is not sensitive to the choice of this factor. Thus, we use
\begin{equation}
\lambda_{E} \approx \gamma \lambda_{F}
\end{equation}
with $\gamma \sim 1$, and below our results are shown for $\gamma = 2$. 

To estimate $\lambda_F$, we employ the single-particle dispersion relation~\cite{tanaka2000} of the 1D Hubbard model in the absence of the electric field $\xi_k = -2t {\rm cos}(ka)-\mu$. Keeping the first two terms in the expansion of $\xi_k$ in $ka$, we obtain 
\begin{equation}
\xi_k \approx \xi_s + \frac{\hbar^2 k^2}{2m_e} -\mu
\end{equation}
with $\xi_s=-2t$ and the effective electron band mass $m_e= \hbar^2/2ta^2$. Then, the Fermi wavenumber is obtained from $\xi_{k_F}=0$ as $k_F=\sqrt{(\mu-\xi_s)/ta^2}$ and the Fermi wavelength $\lambda_F=2\pi/k_F$ is given by \begin{equation}\label{fermilength}
\lambda_F = 2\pi a  \sqrt{\frac{t}{\mu -\xi_s}}=\sqrt{2} \pi a,
\end{equation}
where for the half-filling case we use $\mu=0$.

According to the relation ${\bf E}(x)= -\frac{{\rm d} [V(x)/q]}{{\rm d}x}\hat{\bf{x}}$ with $q=-e$ the electron charge, we obtain the following expression of $V(x)$:
\begin{equation}\label{V}
V(x) = -2\,q\lambda_E\,E_{0}\,e^{-L/2\lambda_{E}} \sinh\big[\big(2x-L\big)/2\lambda_{E}\big].
\end{equation} 
In our calculations, the energy, length and electric field are in units of the hopping rate $t$, the lattice constant $a$ and $t/(ea)$, respectively.

\section{Results and Discussions}\label{sec3}

\subsection{Surface insulating states of superconductors induced by an electric field at $T=0$}
\begin{figure}[htb]
\centering
\includegraphics[width=1\linewidth]{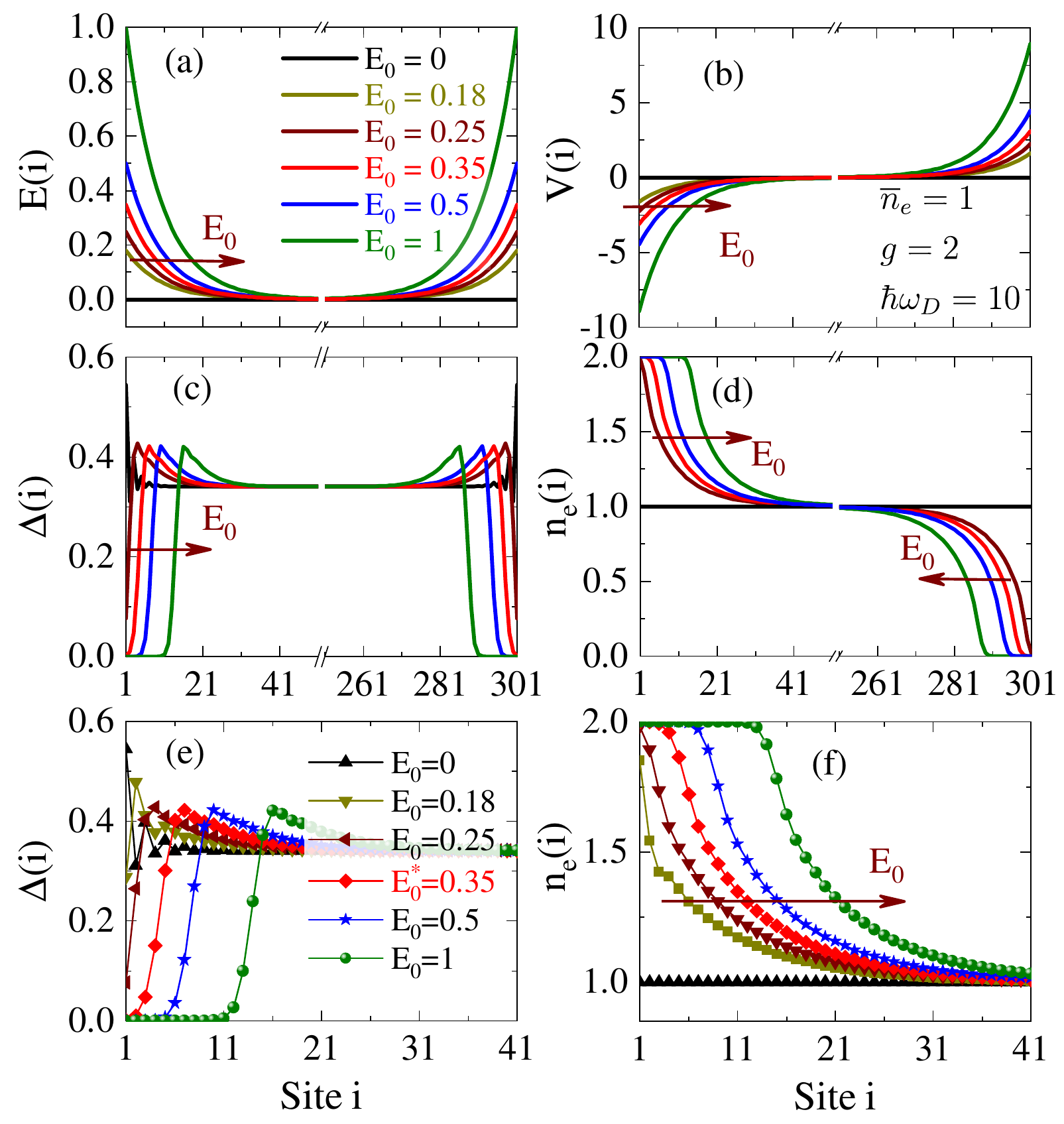}
\caption{(Color online) The screened electric field $E(i)$~(a), the electronic potential energy $V(i)$~(b), the pair potential (order parameter) $\Delta(i)$~(c), and the spatial electron distribution $n_e(i)$~(d) calculated for $E_0=0$, $0.18$, $0.25$, $0.35$, $0.5$ and $1$. Panels (e) and (f) demonstrate zoom-in plots of $\Delta(i)$ and $n_e(i)$ near the left chain edge. The calculations are done at $T=0$ for the material parameters $\bar{n}_e=1$, $g=2$, $\hbar\omega_D=10$, $N=301$ and $\gamma=2$.}
\label{fig1}
\end{figure}

\begin{figure*}[htpb]
\centering
\includegraphics[width=1\linewidth]{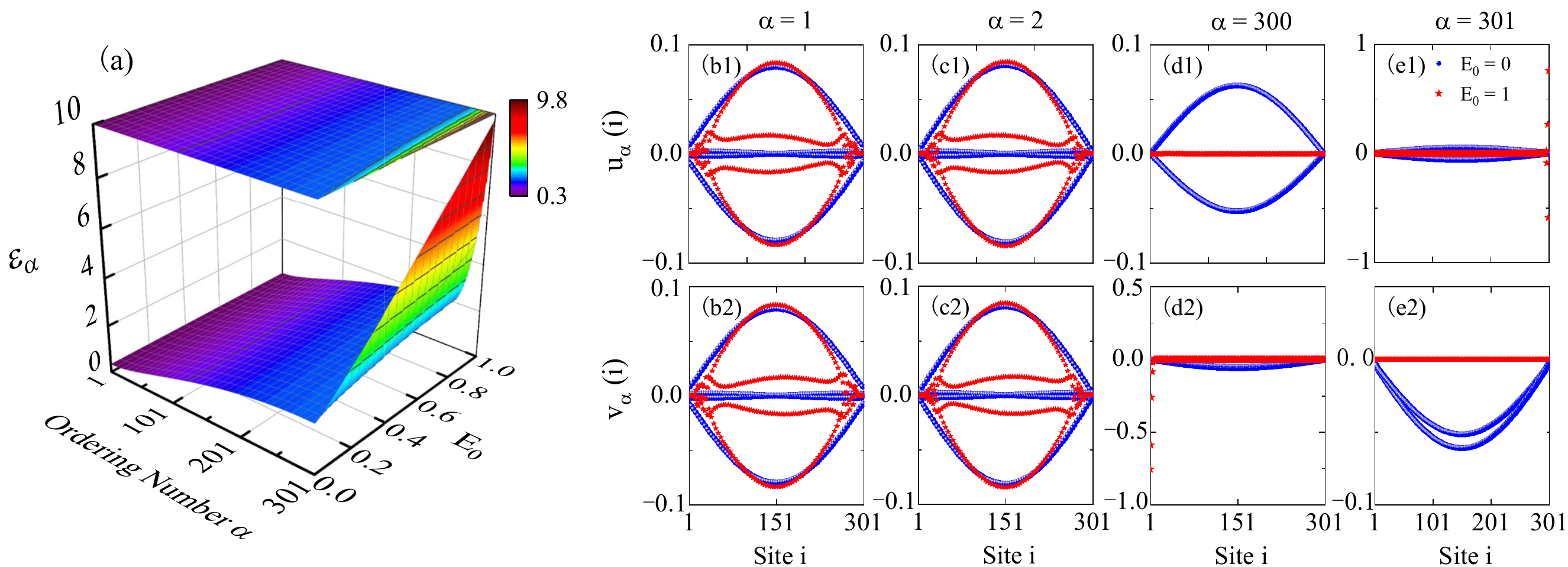}
\caption{(Color online) (a) The quasiparticle energy $\epsilon_\alpha$ as a function of the quantum (ordering) number $\alpha$ and the boundary electric field $E_0$; here the upper panel represents the contour plot of $\epsilon_\alpha$. (b1,b2) The quasiparticle wavefunctions $u_\alpha(i)$ and $v_\alpha(i)$ versus the site number $i$ at $\alpha = 1$;  the same quantities but for $\alpha=2$, $300$ and $301$ are shown in (c1,c2), (d1,d2) and (e1,e2), respectively. In panels (b,c,d,e) the blue points correspond the case of zero field whereas the red ones are for $E_0=1$. The material parameters are the same as in Fig.~\ref{fig1}.}
\label{fig2} 
\end{figure*}

Figure~\ref{fig1} shows a typical example of the surface superconductor-insulator transition induced by a screened electric field $E(i)$ in a one-dimensional superconducting chain with $N = 301$ and $E_0=0,\, 0.18,\, 0.25,\, 0.35,\, 0.5$ and $1$ in the half-filling case at $T=0$. Here the screening length $\lambda_E$ is equal to $9.0$~(as $\lambda_F=4.5$). The coupling constant $g$ is set to $2$ and $\hbar\omega_D=10$.

The electrostatic field $E(i)$ and the corresponding potential energy $V(i)$ are shown in Figs.~\ref{fig1}(a) and (b). As is seen, $E(i)$ drops from $E_0$ at the edges to zero in the center of the chain while $V(i)$ sharply increases with $i$ in the vicinity of the boundaries, according to Eqs.~(\ref{Ei}) and (\ref{V}). The corresponding spatial distribution of the pair potential is given by Fig.~\ref{fig1}(c). One can see that $\Delta(i)$ is nearly uniform in the center of the chain but when approaching an edge (the both left and right), it exhibits a peak with a subsequent abrupt drop to zero. From the zoom-in image in Fig.~\ref{fig1}(e), we learn that the peaks in $\Delta(i)$ shift towards the center of the chain with increasing $E_0$. When $E_0$ crosses the critical value $E_0^*=0.35$, $\Delta(i=1)$ vanishes. Then, this zero-pair-potential region expands with further increasing $E_0$ so that one finds that $\Delta(i\leq11)=0$ for $E_0=1$. The same happens near the opposite edge, where we have $\Delta(i\geq 291)=0$.

Now, we turn to the electron distribution. In the absence of the electric field, $n_e(i)$ is uniform and given by the uniform black line in Fig.~\ref{fig1}(d) and the black line with the up triangles in the zoom-in image of Fig.~\ref{fig1}(f). The character of the distribution changes in the presence of the applied field. Indeed, for $E_0 >0$ one finds that $n_e(i)$ exhibits a significant increase near the left edge and a decrease near the right edge. In the center of the chain $n_e(i)$ approaches the half-filling value. When $E_0$ crosses $E_0^*=0.35$, the site $i=1$ becomes fully occupied (see the red dotted curve), i.e. $n_e(i=1)=2$, which corresponds to the onset of the surface insulating state. At the same time $n_e(i=301)=0$, which also corresponds to the onset of the insulating state at the right edge. Thus, $E_0^*$ can be referred to as the critical electric field of the surface superconductor-insulator transition. For $E_0 > E^*_0$, the surface insulating state expands. For example, we find that $n_e(i\leq 11)=2$ and $n_e(i\geq 291)=0$ for $E_0=1$. Moreover, the surface domains with $\Delta(i)=0$ coincide exactly with the surface insulator domains. Thus, we observe the direct surface superconductor-insulator transition without the presence of an intermediate metallic state.

To go in more detail about the behavior of the pair potential and electron distribution in the vicinity of the surface superconductor-insulator transition, we first investigate the quasiparticle energies $\epsilon_\alpha$ and quasiparticle wavefunctions $u_\alpha(i),\,v_\alpha(i)$, as they are directly related to $n_e(i)$ and $\Delta(i)$ through Eqs.~(\ref{ne}) and (\ref{op}). Figure~\ref{fig2}(a) shows $\epsilon_\alpha$ as a function of $\alpha$ and $E_0$ together with the contour plot of this function. [We recall that $\alpha$ enumerates the quasiparticles states in the energy ascending manner.] The lowest quasiparticle energy in Fig.~\ref{fig2}(a) corresponds to $\alpha=1$ and $E_0=0$~($\epsilon_{\alpha}=0.34$) while the highest one is for $\alpha=301$ and $E_0=1$~($\epsilon_\alpha=9.80$). One can also see that for $\alpha > 250$ the quasiparticle energies notably increase with $E_0$ and, moreover, this increase is much more pronounced for larger $\alpha$. On the contrary, for $\alpha < 250$ the electric-field effect on $\epsilon_\alpha$ is almost negligible. According to Fig.~\ref{fig2}(a), all the quasiparticle states contribute to the pair potential when $E_0\leq1$~[we have $\epsilon_{\alpha} < \hbar\omega_D = 10$, see Eq.~(\ref{op})].

The low-energy and high-energy quasiparticle wavefunctions $u_\alpha(i)$ and $v_\alpha(i)$ with $\alpha=1$, $2$ and $300, 301$, respectively, are illustrated in Fig.~\ref{fig2}(b, c, d, e). The blue dots are the data for $E_0=0$, while the red stars are the results for $E_0=1$. Notice that $u_\alpha(i)$ and $v_\alpha(i)$ are, of course, single-valued functions, and the appearance of different sets of the red and blue data in Fig.~\ref{fig2} is a reflection of fast oscillations of the quasiparticle wave functions from one site to another.

As is mentioned above, the energies $\epsilon_{\alpha=1,2}$ are nearly constant ($\approx 0.34$), when $E_0$ increases from $0$ to $1$. This agrees with the fact that the corresponding quasiparticle wavefunctions are only slightly sensitive to the presence of the electric field. The spatial profiles of $u_{1,2}(i)$ and $v_{1,2}(i)$ for $E_0=0$ are in agreement with the results given in Fig.~4 of Ref.~\cite{croitoru2020}, and are similar to those calculated at $E_0=1$: the maxima of their absolute values are located at $i=151$ while the wavefunctions are almost zero near the boundaries. 

On the contrary, the high-energy quasiparticle wavefunctions with $\alpha=300$ and $301$, are significantly affected by the electric field. For example, this is immediately seen from the data shown in panel (d1). One can also see the presence of significant deviations between the blue ($E_0=0$) and red ($E_0=1$) data near the chain edges in panels (d2) and (e1). These deviations are the signature of the accumulation of charges at the edges of the chain in the presence of a sufficiently strong electric field. For high-energy quasiparticle states with even $\alpha$ we find significant increase of $|v_\alpha(i=1)|$, resulting from the accumulation of electrons at the left edge. For high-energy states with odd $\alpha$ we observe large values of $|u_\alpha(i=301)|$ due to the concentration of positive charges at the right edge of the chain.

\begin{figure}[htpb]
\centering
\includegraphics[width=1\linewidth]{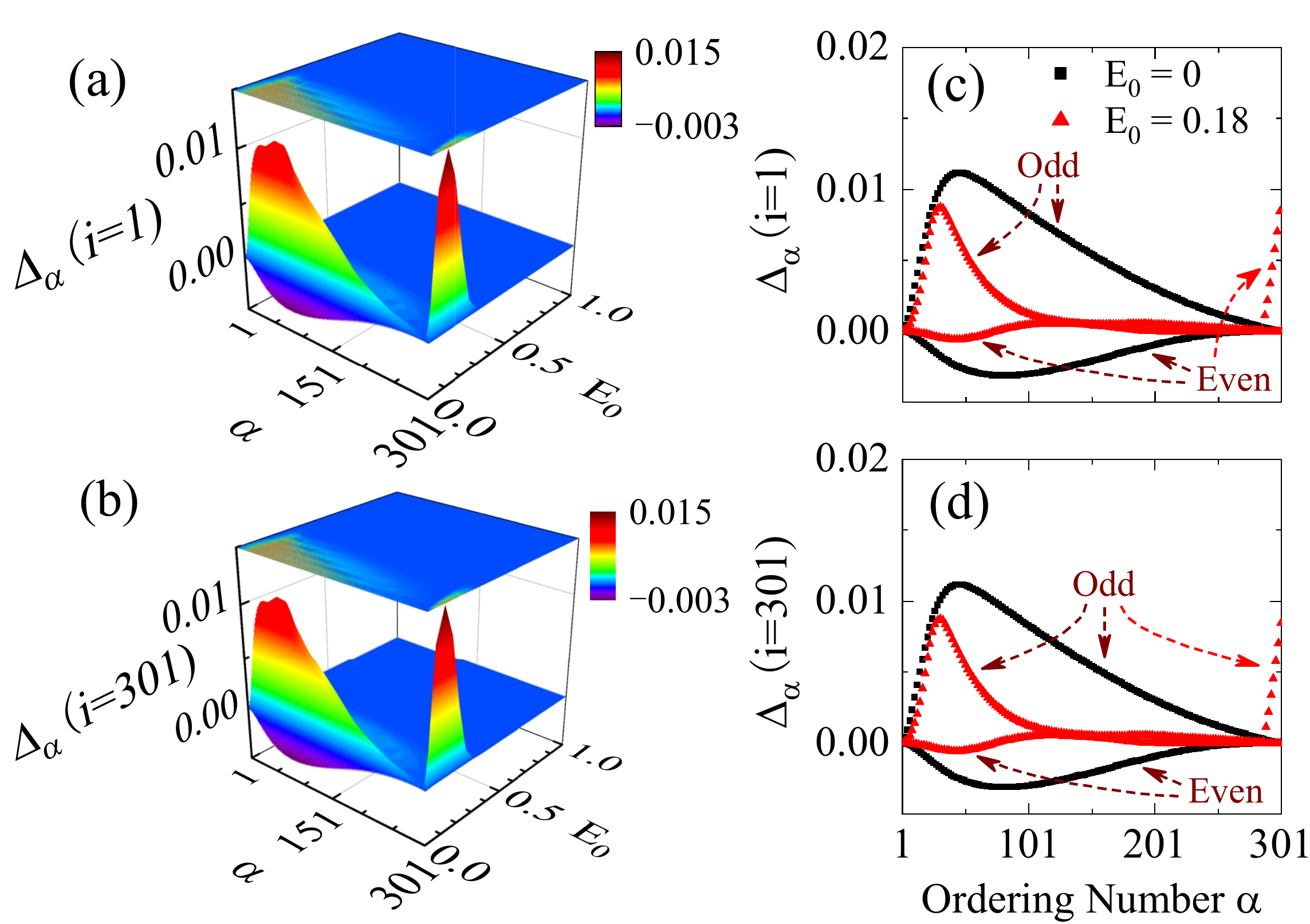}
\caption{(Color online) (a, b) The single-species quasiparticle contribution to the pair potential $\Delta_\alpha(i)$ calculated for $i = 1$ and $301$ and shown as a function of $\alpha$ and $E_0$~(the upper panel is the corresponding counter plot). (c, d) The same quantity but as a function of $\alpha$ at $E_0=0$ (black squares) and $0.18$ (red triangles); here the sets (branches) corresponding to odd and even values of $\alpha$ are displayed. The microscopic parameters are the same as in Fig.~\ref{fig1}.}
\label{fig3}
\end{figure}

Now, we investigate how the quasiparticle properties are connected with changes and suppression of the pair potential near the chain edges. To facilitate our consideration, we introduce 
\begin{equation}\label{op_single}
\Delta_\alpha(i) = g\,u_\alpha(i) v^*_\alpha(i)\big[1-2f_\alpha\big],
\end{equation}
which is the contribution to $\Delta(i)$ of the quasiparticles related to a particular value of $\alpha$. Figure~\ref{fig3}(a, b) demonstrate $\Delta_\alpha(i)$ as a function of $\alpha$ and $E_0$ at the boundaries $i=1$ and $301$, respectively. The upper panels in Figs.~\ref{fig3}(a, b) are the corresponding contour plots. Notice that since $\hbar\omega_D=10$, all quasiparticles with positive energies are inside the Debye window and hence, contribute to the pair potential, as seen from Fig.~\ref{fig2}. The data given in Figs.~\ref{fig3}(a) and (b) look nearly the same but there are minor differences discussed below. For $E_0 \lesssim 0.35$, both $\Delta_\alpha(i=1)$ and $\Delta_\alpha(i=301)$ exhibit two pronounced maxima: one occurs in the domain of low quasiparticle energies while the other (much sharper) takes place about $\alpha\approx301$. The data shown in Figs.~\ref{fig3}(a, b) make it possible to conclude that for $E_0 < 0.35$ the both low- and high-energy quasiparticles make significant contributions to $\Delta(i=1,301)$. However, for $E_0>0.35$ these contributions are significantly depleted, as the blue color in both panels represents nearly zero values of $\Delta_\alpha$.

\begin{figure}[htpb]
\centering
\includegraphics[width=0.8\linewidth]{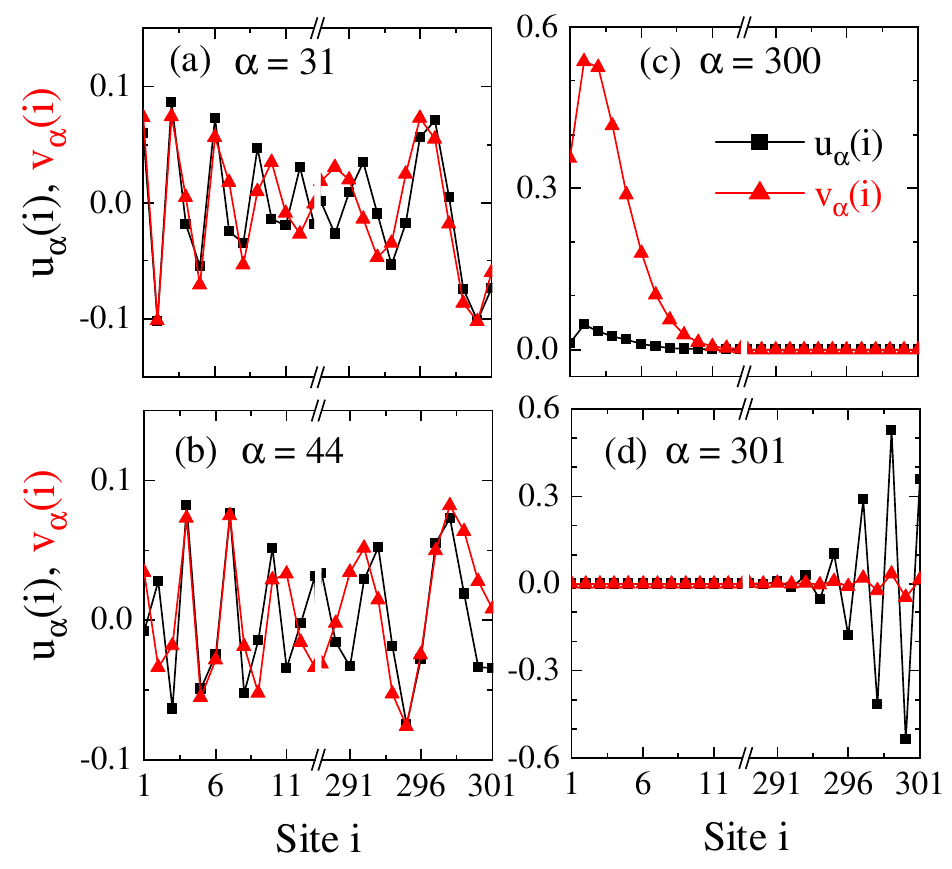}
\caption{(Color online) The quasiparticle wavefunctions $u_\alpha(i)$ and $v_\alpha(i)$ versus the site number $i$ for $\alpha=31$~(a), $44$~(b), $300$~(c) and $301$~(d) at $E_0=0.18$. The material parameters are the same as in Fig.~\ref{fig1}.}
\label{fig4}
\end{figure}

Further details of $\Delta_\alpha(i=1,\,301)$ are given in Figs.~\ref{fig3}(c, d), where $\Delta_\alpha(i=1,\,301)$ are shown as functions of $\alpha$ at $E_0=0$ (black squares) and $E_0=0.18$ (red triangles). First we discuss the results for $E_0=0$. In the absence of the electric field, $\Delta_\alpha(i=1)$ and $\Delta_\alpha(i=301)$ exhibit two branches - positive with odd $\alpha$ and negative with even $\alpha$. The data for $\Delta_\alpha(i=1)$ and $\Delta_\alpha(i=301)$ are the same, which reflects the inversion symmetry of the chain in the absence of the electric field. The fast oscillation between the positive (odd) and negative (even) values of $\Delta_\alpha(i=1,\,301)$ are related to the presence of a $\pi$-phase shift between $u_\alpha(i)$ and $v_\alpha(i)$ at the boundaries $i=1$ and $i=301$~[see Eq.~(\ref{op_single}) at $T=0$]. In particular, $u_\alpha(i=1,301)$ and $v_\alpha(i=1,301)$ have the same sign for a quasiparticle with odd $\alpha$ while they have opposite signs for even $\alpha$. Similar results can be seen from Figs.~\ref{fig4}(a) and (b), where $u_\alpha(i)$ and $v_\alpha(i)$ are shown for $\alpha=31$ and $44$~[here the data are for $E_0=0.18$]. The maximal contribution of the positive branch for $E_0=0$ occurs at $\alpha=47$ while the most pronounced but less significant (as compared to the odd states) input of the negative branch is at $\alpha=82$, which matches the slope variation of the accumulative pair potential in Fig.~3 of Ref.~\cite{bai2023}.

Now, let us consider $\Delta_\alpha(i=1,\,301)$ calculated for $E_0=0.18$ and also given in Figs.~\ref{fig3}(c) and (d). Similarly to the data for zero field, we again have positive and negative branches in the dependence of $\Delta_\alpha(i=1,\,301)$ on $\alpha$. As compared to the case of $E_0=0$, the low-energy maximum and minimum of these two branches become less pronounced, reflecting the appearance of the additional local maximum of the positive branch due to the high-energy quasiparticles. The positions of these low-energy minimum and maximum are shifted towards smaller values of $\alpha$, i.e. to $\alpha=31$ and $\alpha=44$, respectively. For $\alpha < 150$ the positive and negative branches still correspond to odd and even $\alpha$, which is the same as in the case of $E_0=0$. However, this correspondence is broken for high quasiparticle energies. In particular, the situation changes dramatically for $\alpha>285$. Here the positive branch for $i=1$ correspond to even $\alpha$ values [see Fig.~\ref{fig3}(c)], while the positive branch for $i=301$ is related to odd $\alpha$~[see Fig.~\ref{fig3}(d)]. This is dictated by the breakdown of the inversion symmetry due to the presence of the electric field. For example, as shown in Fig.~\ref{fig4}(c), $u_\alpha(i=1)$ and $v_\alpha(i=1)$ for $\alpha=300$ are finite and positive at $i=1$ while both wavefunctions are nearly zero at the other boundary $i=301$. It means that the contribution of the states with $\alpha=300$ to the pair potential at $i=301$ is nearly zero. However, the quasiparticles with $\alpha=301$ are accumulated near $i=301$ so that their contribution to the order parameter is depleted at $i=1$. Thus, when the electric field is switched on, we find complex rearrangement of the quasiparticle spatial distributions, and this is related to significant depletion of the pair potential near the chain edges.

\begin{figure}[htbp]
\centering
\includegraphics[width=1\linewidth]{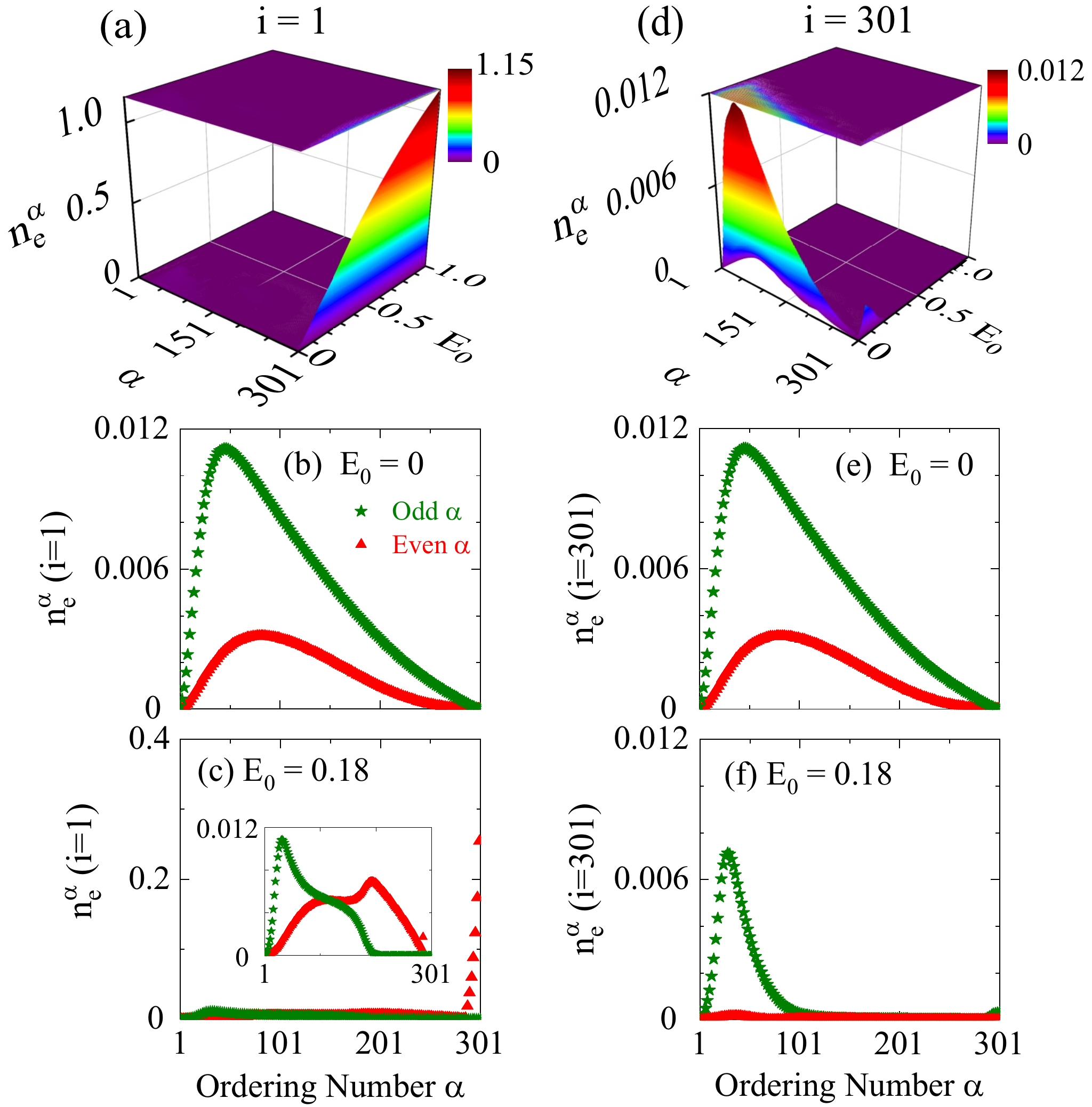}
\caption{(Color online) (a,d) The single-species quasiparticle contribution to the electron density $n_e^\alpha$ calculated for $i=1$ and $301$, respectively, and shown as a function of $\alpha$ and $E_0$; (b,e) $n_e^\alpha(i=1,301)$ as a function of $\alpha$ for $E_0=0$; (c,f) $n_e^\alpha(i=1,301)$ versus $\alpha$ for $E_0=0.18$. The green stars represent odd values of $\alpha$ whereas the red triangles are the data for even $\alpha$. The material parameters are the same as in Fig.~\ref{fig1}.}
\label{fig5}
\end{figure}

To explore the accumulation/depletion of the charge carriers at the chain edges, we consider 
\begin{equation}\label{nea}
n_e^{\alpha}(i) = 2 \bigg\{ f(\epsilon_\alpha)|u_\alpha(i)|^2 +  \big[1-f(\epsilon_\alpha)\big]|v_\alpha(i)|^2\bigg\},
\end{equation}
which is the contribution of the quasiparticles with the quantum number $\alpha$ to $n_e(i)$. In Figs.~\ref{fig5}(a, d), $n_e^{\alpha}(i)$ is shown as a function of $\alpha$ and $E_0$ at $i=1,301$. For more detail, Figs.~\ref{fig5}(b, c, e, f) demonstrate $n_e^{\alpha}(i)$ as a function of $\alpha$, calculated for $i=1$ and $i=301$ at $E_0=0,0.18$. The inset in Fig.~\ref{fig5}(c) is the zoom-in plot. The contributions of quasiparticles with odd and even $\alpha$ are given by green stars and red triangles, respectively.

From Fig.~\ref{fig5}(a), one can see that $n_e^\alpha(i=1)$ increase significantly with $E_0$ for high-energy quasiparticles with $\alpha>280$. However, only low energy quasiparticles with $\alpha\approx 50$ contribute to $n_e(i=301)$. Furthermore, this contribution is notable only at the fields with $E_0<0.35$. When $E_0$ exceeds $0.35$, all quasiparticles produce zero contribution to $n_e(i=301)$.

As seen from Fig.~\ref{fig5}(b, e), the profiles of $n_e^\alpha(i=1,\,301)$ are the same in the absence for zero field (the inversion symmetry). At $E_0 =0.18$, $n_e^\alpha(i=1)$ for $i=1$ differs significantly from that for $i=301$. In particular, when $E_0$ increases from $0$ to $0.18$, the odd-$\alpha$ branch of $n_e^{\alpha}(i=301)$ decreases significantly so that its maximum drops from $0.011$ to $0.007$. At the same time the even-$\alpha$ branch of $n_e^{\alpha}(i=301)$ nearly approaches zero. It means that $n_e(i=301)$ exhibits a notable decrease due to the presence of an applied electric field, and we have the concentration of the positive charge near the right edge of the chain. For $n_e^{\alpha}(i=1)$ one finds a qualitatively different picture. Though the contributions of the quasiparticles with $\alpha < 280$ decrease with increasing $E_0$, the sector of high-energy states exhibits a huge increase of $n_e^{\alpha}(i=1)$. As a result, $n_e(i=1)$ increases significantly when $E_0$ rises from $0$ to $0.18$, which is clearly the reflection of the electron accumulation near the left edge of the chain due to the applied electric field.

\subsection{Phase diagram of surface insulating states}
\begin{figure}[htpb]
\centering
\includegraphics[width=0.6 \linewidth]{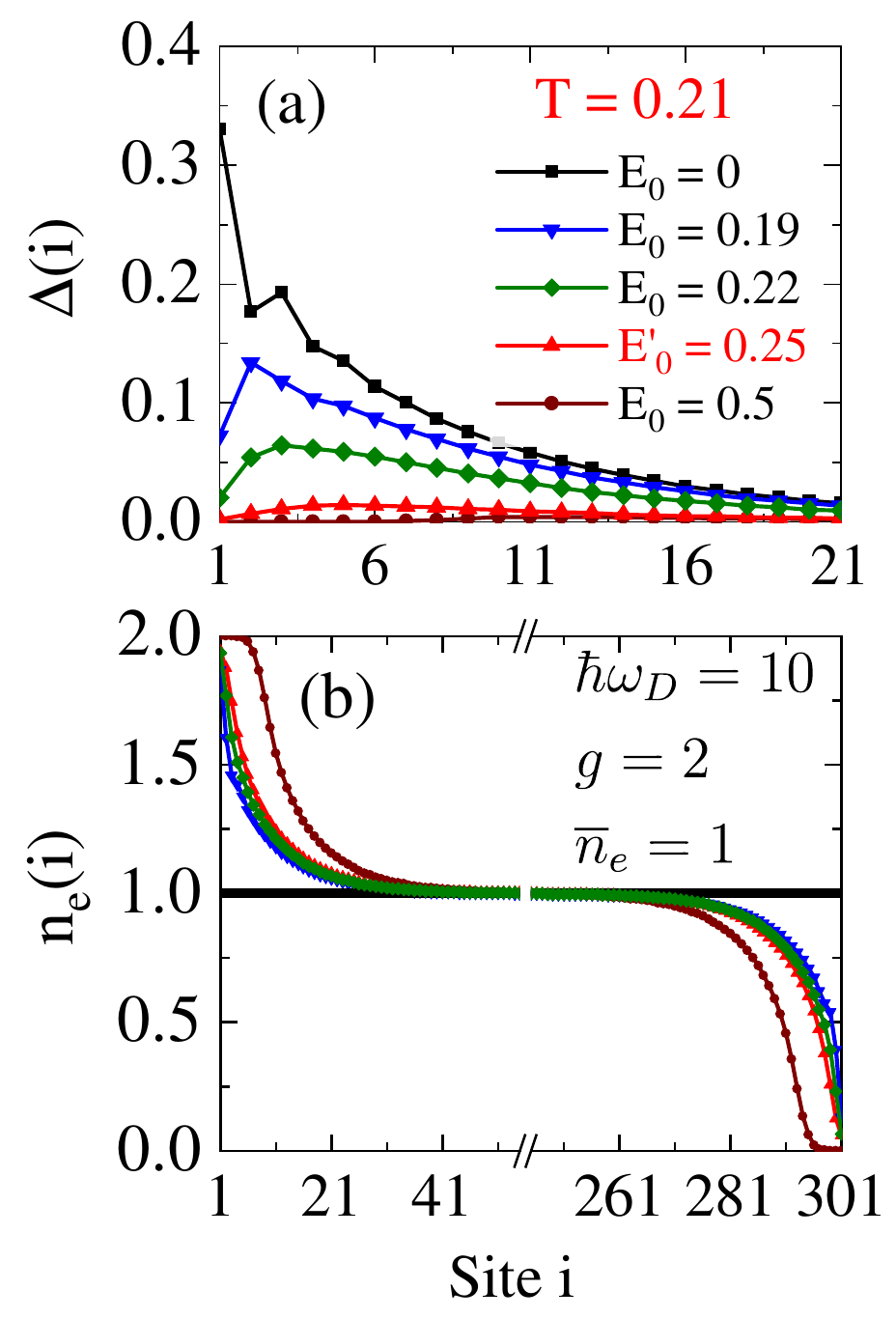}
\caption{(Color online) The spatial pair potential $\Delta(i)$ and election distribution $n_e(i)$ at $T=0.21$ with $E_0=0$, $0.19$, $0.22$, $0.25$ and $0.5$. The other parameters are set as the same in Fig.~\ref{fig1}. The boundary electric field $E'_0$ corresponds to the situation $\Delta(i=1,301)=0$.}
\label{fig6}
\end{figure}

Here we study the phase diagram of the surface superconducting, metal (normal), and insulating states as dependent on the temperature $T$ and external field $E_0$. To have an idea about the temperature effect on the superconductor-insulator transition, $\Delta(i)$ and $n_e(i)$ are shown in Fig.~\ref{fig6} for $E_0=0$, $0.19$, $0.22$, $0.25$ and $0.5$ at $T=0.21$. The other parameters of the calculation are the same as in Fig.~\ref{fig1}. As is seen, $\Delta(i=1)$~[an also $\Delta(i=301)$] becomes zero when $E_0$ crosses the value $E'_0=0.25$, which differs significantly from $E_0^*=0.35$ at $T=0$~[c.f. Fig.~\ref{fig1}]. At the same time we find that $n_e(i=1) = 1.94$~[while $n_e(i=301)=0.06$] at $E_0=E'_0$. It means that there is no full occupation for $i=1$ at $E_0=E'_0$, and also, the site $i=301$ is not completely empty in this case. Thus, $E'_0$ marks the onset of the surface normal state rather than the insulating one. The surface insulating state appears at $T=0.21$ only when $E_0$ crosses the critical value $E^*_0=0.44$. For larger fields $n_e(i=1) = 2$ and $n_e(i=301)=0$, as seen in Fig.~\ref{fig4}. Thus, at finite temperatures the electric-field-induced superconductor-insulator transition is replaced by the superconductor-metal-insulator transition. When $E_0$ increases at $T=0.21$, one first finds the superconducting-normal transition at $E_0=E'_0=0.25$ and then, the metal-insulator transition at $E_0=E^*_0=0.44$.   
\begin{figure}
\centering
\includegraphics[width=0.9\linewidth]{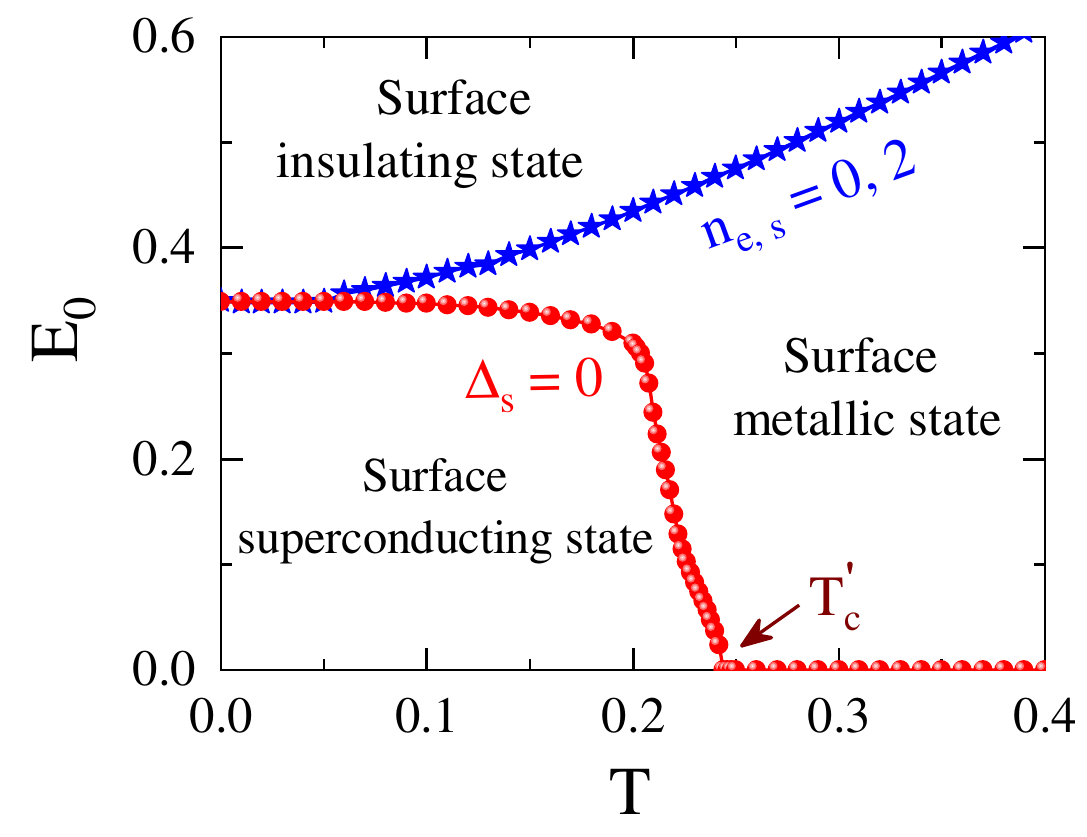}
\caption{(Color online) The phase diagram of the surface superconductor-metal-insulator transition in $E_0$-$T$ plane. The superconducting-normal boundary is given by the curve above which $\Delta_s=\Delta(i=1,301)=0$. The metal-insulator boundary marks the onset of the insulating state $n_e(i=1)=2$ and $n_e(i=301)=0$, which is refereed to as $n_{e,s}=0,2$.} 
\label{fig7}
\end{figure}

Figure~\ref{fig7} shows the phase diagram in the $E_0$-$T$ plane that describes the surface superconductor-metal(normal)-insulator states calculated for the 1D chain. All the material parameters are the same as in Fig.~\ref{fig1}. The red solid circles represent the boundary between the surface superconducting and normal states above which $\Delta_s=\Delta(i=1,301)=0$. The blues stars yield the boundary between the metallic and insulating states. Above this boundary we have $n_e(i=1)=2$ and $n_e(i=301)=0$. Below the lower boundary the surface of the sample is superconducting. Above the upper boundary we have the surface insulating state. Between the boundaries the surface of the system is in the normal metallic state.

As is seen, when the temperature increases, the critical value $E'_0$ remains nearly the same up to $T=0.2$. Then, it drops rapidly and becomes zero at $T'_{c}=0.244$. On the other hand, $E^*_0$, that marks the onset of the surface insulating state, slowly increases with the temperature from $0.35$ to $0.6$ when the temperature goes from $0$ to $0.4$. This increase is due to the thermal smearing in the Fermi-Dirac distribution. One finds that the lower and upper boundaries approach each other at $E_0\approx 0.35$ for $T<0.05$~($\approx20.5\%T'_{c}$). Thus, one can expect that the direct superconductor-insulator transition occurs at $T<0.05$. 
We remark that surprisingly, the qualitative picture of our results for the surface transformation under the applied electric field is in qualitative agreement with the phase diagram of the superconductor-insulator transition in (Li, Fe)OHFeSe thin flakes obtained by the transport measurements~\cite{ma2019, yin2020}. This is especially true of the phase boundary between the superconducting and insulating states at low temperatures.

Finally, based on our results, we can estimate the strength of $E_0$. The data shown in Fig.~\ref{fig1}(c) demonstrate that $\Delta_b \approx t/3 < \mu_F-\xi_s=2t$, with $\Delta_b$ the bulk pair potential at $i=151$, which is beyond the strong-coupling limit (i.e. $\Delta \geq \mu_F-\xi_s$)~\cite{okazaki2014}. As an example of the weak-coupling superconductor, we can use SrTiO$_3$. For this material we have $\Delta\approx0.1$ meV~\cite{ahadi2019} and the averaged lattice constant $a\approx4\AA$~\cite{ohama1984}. Then, one finds $t/(ea)=7.5\times10^5$ V/m, where the above relation between $\Delta_b$ and $t$ is utilized. Therefore, the transition electric field from the surface superconducting state to the surface insulating state at $T=0$ is estimated as $E_0^*=0.35\,t/(ea) = 2.6 \times10^5$ V/m, which is two orders of magnitude lower than the dielectric breakdown field ($3.1\times10^7$ V/m) of SrTiO$_3$ film~\cite{wang2001}. For superconductors with $\Delta=10$ meV, keeping $a\approx4\AA$ and the same relation between $\Delta_b$ and $t$, we have $E_0^*=2.6 \times10^7$ V/m. This electric field is also available in experiments~\cite{paolucci2019}. Even the fields of the order of $10^{10}$ V/m can be achieved based on the voltage-induced polarization of an electrolyte~\cite{dhoot2006}.

\section{Conclusions}\label{sec4}

In conclusion, the electric-field-induced surface insulating state is revealed in a superconductor by numerically solving the Bogoliubov-de Gennes equations for the one-dimensional attractive Hubbard model in a self-consistent manner. We find that the surface insulating state appears once the chain sites near the edges are either fully occupied by electrons or completely empty. This rearrangement occurs due to the applied electric field, affecting the electron distribution near the surface and suppressing the surface pair potential. At zero temperature we find the superconductor-insulator phase transition arising when increasing the electric field. At finite temperatures the system first undergoes the surface superconductor-metal transition and then, at larger fields, the metal-insulator phase transition. The phase diagram of the surface superconducting, metallic, and insulating states is obtained for a wide range of the temperatures and applied electric fields. Remarkably, this diagram qualitatively matches the results of the transport measurements in (Li, Fe)OHFeSe thin flakes~\cite{ma2019, yin2020}.

\begin{acknowledgments}
This work was supported by Science Foundation of Zhejiang Sci-Tech University(ZSTU) (Grants No. 19062463-Y \& 22062336-Y), Open Foundation of Key Laboratory of Optical Field Manipulation of Zhejiang Province (ZJOFM-2020-007). The study has also been funded within the framework of the HSE University Basic Research Program.
\end{acknowledgments}


\end{document}